\documentclass[twocolumn,preprintnumbers,showpacs,prb]{revtex4-1}
\usepackage{graphicx}
\usepackage{dcolumn}
\usepackage{multirow}
\usepackage{epsfig}
\begin{document}
\title{High-Temperature Phonon Spectra of Multiferroic BiFeO$_3$ from Inelastic Neutron Spectroscopy}
\author{Mohamed Zbiri}
\email{Electronic mail: zbiri@ill.fr.}
\affiliation{Institut Laue-Langevin, BP 156, F-38042 Grenoble Cedex 9, France}
\author{Helmut Schober}
\affiliation{Institut Laue-Langevin, BP 156, F-38042 Grenoble Cedex 9, France, and,\\
Universit\'e Joseph Fourier, Unit\'e de Formation et de Recherche (UFR) de Physique, F-38041 Grenoble Cedex 9, France}
\author{Narayani Choudhury}
\affiliation{Department of Physics, University of Arkansas, Fayetteville, Arkansas 72701, USA}
\author{Ranjan Mittal, Samrath L. Chaplot}
\affiliation{Solid State Physics Division, Bhabha Atomic Research Centre, Trombay, Mumbai 400 085, India}
\author{Sadequa J. Patwe, Srungarpu N. Achary, Avesh K. Tyagi}
\affiliation{Chemistry Division, Bhabha Atomic Research Centre, Trombay, Mumbai 400 085, India}
\begin{abstract}
We report inelastic neutron scattering measurements of the phonon spectra in a pure powder sample of the multiferroic material BiFeO$_3$. A high-temperature range was covered to unravel the changes in the phonon dynamics across the N\'eel (T$_N$ $\sim$ 650 K) and Curie (T$_C$ $\sim$ 1100 K) temperatures. Experimental results are accompanied by ab-initio lattice dynamical simulations of phonon density of states to enable microscopic interpretations of the observed data. The calculations reproduce well the observed vibrational features and provide the partial atomic vibrational components. Our results reveal clearly the signature of three different phase transitions both in the diffraction patterns and phonon spectra. The phonon modes are found to be most affected by the transition at the T$_C$. The spectroscopic evidence for the existence of a different structural modification just below the decomposition limit (T$_D$ $\sim$ 1240 K) is unambiguous indicating strong structural changes that may be related to oxygen vacancies and concomitant Fe$^{3+}$ $\rightarrow$ Fe$^{2+}$ reduction and spin transition.
\end{abstract}
\maketitle

Multiferroism is the combination of two or more of the following properties: ferroelectricity, ferromagnetism and ferroelasticity. In particular, tuning the electric or magnetic 
properties by applying a magnetic or an electric field, respectively, is at the basis of new applications in microelectronics, spintronics, data storage and computing 
hardware~\cite{bfo1,bfo2,revphysandapps}. This explains the strong interest in multiferroic materials from a practical point of view. They unfortunately do not abound in nature and are 
often difficult to synthesize technologically. Within the most important ABO$_3$ perovskite family the scarcity of multiferroic properties is mainly due to the $d$-electrons of the B 
cation, which on one side are necessary for magnetism  but on the other side tend to reduce the tendency for an off-centering ferroelectric distortion~\cite{bfo3}. A remarkable exception 
is bismuth ferrite BiFeO$_3$ (BFO). This material seems to be unique in exhibiting multiferroic behavior above room temperature. It is simultaneously ferroelectric, antiferromagnetic and 
ferroelastic, and is subject to crystallographic distortions. The mixture of these multiple order parameters involved in its phase transitions is believed to lead to a very rich phase diagram~\cite{redfern,revphysandapps}. The physical behavior of BFO is characterized by two largely different transition temperatures; antiferromagnetic with T$_N$ $\sim$ 600 - 650~K and ferroelectric with T$_C$ $\sim$ 1050 - 1100~K, and a polarization
that can reach a very large~\cite{revphysandapps,polarization} value $\sim$ 100 $\mu$C/m$^2$. The large value of the Curie temperature could be at the origin
of the observed colossal spontaneous  polarization through a strong ferroelectric ordering~\cite{bfo4}. The polarization in BFO is mainly caused
by the Bi $6s^2$ lone pair, so that ferroelectricity is  induced primarily via this site (strong cation 
displacements). Another remarkable aspect of BFO is the occurrence of FeO$_6$ octahedra tilts accompanying the strong Bi$^{3+}$ displacements at room temperature~\cite{hill1,hill2}. Below 
T$_N$ the magnetic ordering is non-collinear with a G-type-like cycloidal spin-spiral arrangement and the structure is incommensurate~\cite{period} with an extremely long period of 
620~$\AA$. The coexistence of high-temperature magnetism and strong ferroelectricity makes BFO a material with physical properties of practical importance~\cite{revphysandapps,polarization}. Due to these 
exceptional characteristics, BFO has become one of the most studied multiferroic materials leading to a wealth of experimental and simulation results covering 
both structure and dynamics ~\cite{revphysandapps,finittemp,calcmagnetoelec,irphonons,phasestabstructtemp,ptphase,massa,stroppa}. The keen and continuous interest is equally motivated by the 
controversies raised by a priori incompatible experimental findings: determination and interpretation of phase diagram, occurence and nature of phase transitions, assignment and
characterization of frequency modes from different spectroscopy techniques
(Raman, infrared, ...)~\cite{revphysandapps,contr1,contr2,phasestabstructtemp,massa}.

Investigations of the vibrational dynamics could help to lift at least some of the controversies. Numerous Raman and infrared measurements~\cite{bfo5,irphonons,revphysandapps,massa,phasestabstructtemp} show that the different phase transitions leave clear signatures in the vibrational modes. The limitation of optical spectroscopy to the Brillouin zone centre, however, prevents 
obtaining a complete picture of the interplay of dynamics and structural instabilities. In particular zone boundary modes could contribute to the unusual properties of BFO like the 
non-centrosymetric positional shift, which induces ferroelectricity~\cite{bfo5,bfo6}. The strong debate concerning the mode assignments from Raman and infrared spectroscopies~\cite{irphonons,massa,raman1,raman2} just reflects the difficulties in establishing the structural phase diagram. It thus seems obvious that the structural information obtained 
from different diffraction techniques should be adequately complemented by spectroscopic investigations covering the full  Brillouin zone~\cite{phasestabstructtemp,pnd1,pnd2,pnd3}. A 
concrete argument in favor of adopting such techniques, like inelastic neutron scattering (INS), arises from the fact that diffraction experiments allow to capture only the time-averaged 
structure with the consequence of missing subtle structural instabilities and distortions that pervoskites are often subjected to. At complete odds with the importance of BFO, no INS measurements across 
both N\'eel and Curie transition temperatures have been attempted up to now. The present letter aims at closing this gap by probing 
phonon dynamics over the entire Brillouin zone and over a wide high-temperature range covering both the T$_N$ and T$_C$. We devote particular attention to the crossing of the Curie temperature T$_C$ where the ferroelectric-to-paraelectric transition occurs and concomitantly the electric polarization should show a sudden drop. Furthermore, within this high temperature domain some important changes in the 
electronic properties occur like the observed metal-to-insulator transition around 1200~K and a presumably high-spin to low-spin transition~\cite{revphysandapps,calcmagnetoelec,pnd1,pnd2,massa}, resulting from the reduction of Fe$^{3+}$ to Fe$^{2+}$. The latter, if it involves a large fraction of the ions, can lead to an increase in the cell volume and, therefore, should 
be accompanied by a strong dynamical signature.

Polycrystalline samples of BiFeO$_3$ were prepared by solid state reaction of stoichiometric amounts of Bi$_2$O$_3$ (Alfa, 99.5 \%) and Fe$_2$O$_3$ (Alfa, 99.5 \%). Bi$_2$O$_3$ was heated 
to 800 $^{\circ}$C in order to avoid any carbonate or oxycarbonate in the starting material. Fe$_2$O$_3$ was heated to around 300 $^{\circ}$C in order to eliminate any adsorbed water 
molecules. Appropriate amounts of Bi$_2$O$_3$ and Fe$_2$O$_3$ were thoroughly homogenized using agate mortar pestle with acetone as grinding medium. The well homogenized powder sample 
was pelletized to a disk of about 1'' diameter and 0.5'' thickness and heated to 100-150 $^{\circ}$C to remove any residual organic component. Further the pellet was inserted directly 
into a furnace with a preset temperature of 700$^{\circ}$C and soaked for about 4~h and then quenched to ambient temperature by removing the sample from the hot zone of the furnace. At 
this stage the majority of sample formed as rhombohedral BiFeO$_3$. Further the pellet was crushed to powder and repelletized and heated in a similar manner at 850 $^{\circ}$C for 4~h and 
quenched to ambient temperature. The sample obtained after this heat treatment was characterized by XRD recorded on a Philips Xpert-pro diffractometer in the anglular range of 
10-110$^{\circ}$ with a step width of 0.02$^{\circ}$ using Ni filtered Cu-K radiation. All the observed reflections could be assigned to BiFeO$_3$ (rhombohedral R3c). The powder XRD data 
was further analyzed by Rietveld refinement with the reported structural data for BiFeO$_3$. The final refined unit cell parameters a = 5.5775(2) and c = 13.8648(5)~\AA\ are in good 
agreement with parameters reported earlier for BiFeO$_3$.

The INS measurements were performed at the Institut Laue Langevin (ILL) (Grenoble, France) on 8 grams of powder sample prepared as described above. We used the cold neutron 
time-of-flight spectrometer IN6 operating with an incident wavelength $\lambda_{i}$~= 4.14 {\AA} (E$_i$=4.77 meV), which provides a good resolution within the considered dynamical 
range for the anti-Stokes side. The data analysis was done using ILL software tools and the dynamical structure factor $S(Q,E)$, diffractograms and the generalized density of 
states (GDOS) were evaluated using standard ILL procedures without applying multiphonon corrections.

We complement the experiment with density functional based ab-initio lattice dynamical calculations to (i) reproduce the observed generalized phonon density of states (GDOS) and (ii) to extract 
the partial (atomic) vibrational components contributing to the total spectrum. The starting geometry for the calculations was the experimentally refined structure 
in the rhombohedral phase R3c space group. A G-type magnetic ordering was assumed as a spin configuration to account for magnetism of Fe cations. 
Relaxed geometries, total energies and phonon frequencies were obtained within the 0 K first principles framework (which allows a correct treatment of the ground state), using similar 
computational procedure as described previously~\cite{sgo}. In order to determine accurately all the force constants, the supercell approach was used
for lattice dynamics calculations.  A supercell (8$\times$a, 8$\times$b, 8$\times$c) was constructed from the relaxed geometry containing 16 formula units (80 atoms). Total energies and Hellmann–Feynman 
forces were calculated for 18 structures resulting from individual displacements of the symmetry inequivalent atoms in the supercell, along with the inequivalent cartesian
directions $\pm$x, $\pm$y, and $\pm$z.

\begin{figure*}
\includegraphics[angle=270.0,width=18cm]{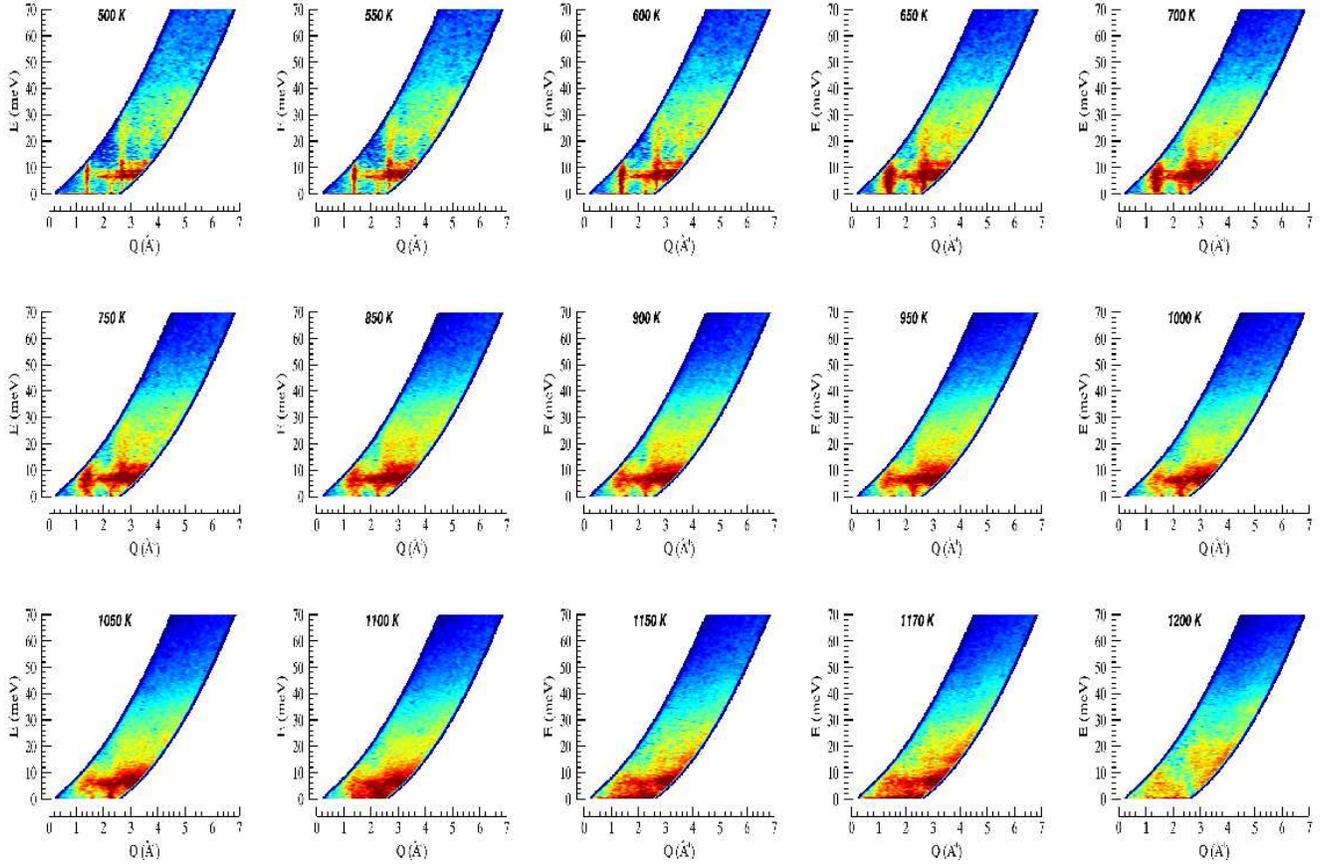}
\caption{(Color online) Temperature-dependent evolution of the experimental Bose-factor corrected dynamical structure factor $S(Q,E)$ of BFO. For clarity, a logarithmic 
representation is used for the intensities; dark red and dark blue refers to strong and weak amplitudes, respectively.}
\label{sqw}
\end{figure*}

\begin{figure*}
\includegraphics[angle=270.0,width=19cm]{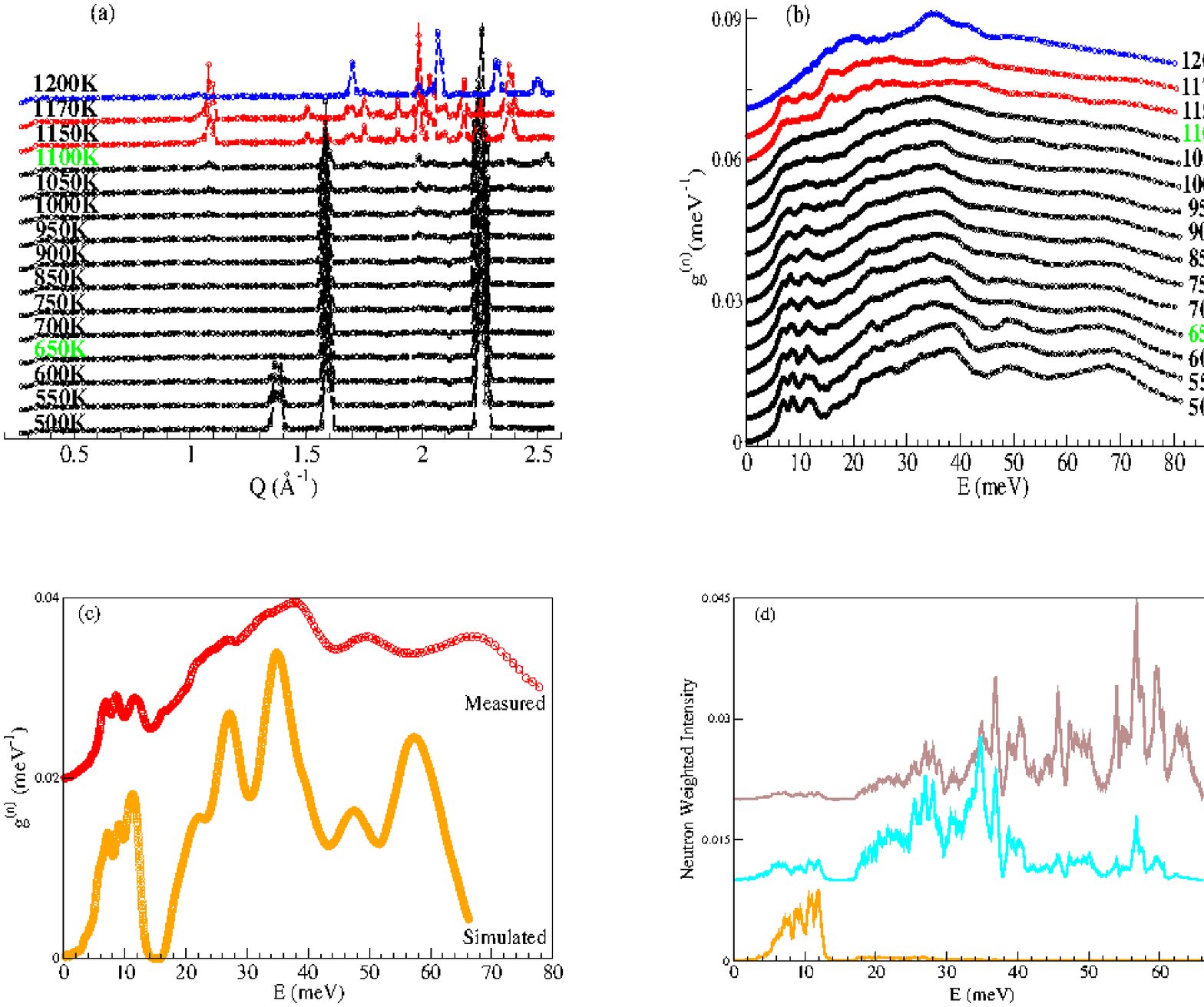}
\caption{(Color online) Temperature dependence of diffraction patterns (a) and phonon generalized density of states (GDOS) spectra (b) of BFO extracted
  from time-of-flight measurements on IN6. Simulated total GDOS and the related partial atomic contributions from {\it ab-initio} calculations are shown in (c) and (d), respectively. On the same plot (c) the calculated GDOS is compared with a measured spectrum. For the sake of clarity diffractograms and all spectra in (a), and (b), (c) and (d), respectively, are vertically shifted with respect to each other. For a better visibility, stability regions and characteristic temperatures T$_N$ and T$_C$ are highlighted with different colors on (a) and (b).}
\label{diffgdos}
\end{figure*}

The temperature evolution~\cite{furnace} of the dynamic structure factor $S(Q,E)$ (Figure~\ref{sqw}) reveals considerable changes up to 1200~K. As a first step it is useful to concentrate on the 
elastic response in order to verify the phase properties. The diffraction patterns obtained from the time-of-flight data are reported on Figure~\ref{diffgdos}(a). As indicated by the 
complete disappearance of the magnetic Bragg peak at 1.4~\AA$^{-1}$ the system loses magnetic ordering close to 650~K. As the temperature increases further the intensities of the two 
Bragg features located at 1.6 and 2.3~\AA$^{-1}$ decrease steadily reflecting the increased thermal motion of the ions. At T$_C$ (1050 - 1100 K)  the diffraction pattern changes 
dramatically indicating that a structural phase transition took place. As the system is heated further there is a second abrupt change of the diffraction pattern at about 1200~K, i.e. 
slightly below the decomposition temperature T$_D=1240$~K. These findings are consistent with recent powder neutron diffraction studies~\cite{pnd1} reporting the probable existence of 
three phases $\alpha$ (T$_\alpha$$<$ T$_C$), $\beta$ (T$_C$$<$T$_\beta$$<$T$_\gamma$) and $\gamma$ (T$_\beta$$<$T$_\gamma$$<$T$_D$).

In view of the strong debate concerning the existence of these phases in the bulk material BFO, Figure~\ref{diffgdos}(b) shows the phonon spectra, in terms of generalized density of 
states (GDOS) over the explored temperature range~\cite{furnace}. In order to analyze the temperature-dependent dynamic response contained in the measured GDOS we use the 
ab-initio simulated spectrum as a reference point. Figure~\ref{diffgdos}(c) compares calculated and observed phonon spectra in the ferroelectric $\alpha$ phase. In order to compare with 
experimental data, the calculated GDOS was determined as the sum of the partial vibrational densities of states $g_{i}$ weighted by the atomic scattering cross sections (as seen by 
neutrons) and masses: GDOS=$\sum_{i} (\sigma_i/M_i)g{_i}$, where ($\sigma_i/M_i$ = 0.264 (O), 0.207 (Fe), 0.044 (Bi); i=\{O, Fe, Bi\}). All the observed features are well reproduced by 
the calculations. The simulated phonon frequencies and intensity ratios between the different peaks are in good agreement with the measurements. Based on this we can consider that the 
experimental data validate the computational approach. We notice that our calculated GDOS matches also other available theoretical work~\cite{calcGDOS}. A more detailed look at the 
calculated phonons of BFO is shown in Figure~\ref{diffgdos}(d). Therein the partial phonon density of states (PDOS) are represented. There is a contribution from all the atoms in the 
range 0 - 14~meV with an increasing amplitude as the atomic mass increases, i.e, Bi has the highest intensity in this region. The low-frequency region is separated from the rest of the 
spectrum by a gap (14-17~meV). Beyond 14 meV the Bi cations do not exhibit any dynamics. The Fe and O density of states are both strongly present in the mid-part of the spectrum spreading 
from 14~$<$E$<$~45~meV. The oxygens stretch modes dominate  at the highest frequency end of the spectrum (45 $<$E$<$ 70 meV). It should be noted that available high-temperature optical 
spectra~\cite{massa} shows a similar trend comparing to the presently reported phonon spectra in terms of temperature-dependent dynamical evolution of BiFeO$_3$.

The phonon dynamics as a function of temperature consistently reflects the suite of phase transitions deduced from the diffraction data. Apart from some indication for a broadening 
of the spectra in the oxygen dominated region beyond 45~meV the magnetic transition leaves a rather weak trace in the phonon spectrum. It is mainly characterized by the appearance of 
strong paramagnetic scattering at the location of the magnetic Bragg peaks when the transition is passed (Figure~\ref{sqw}). The broadening of the phonon bands and concomitant filling of 
the gap continuos when heating within the polar ferroelectric $\alpha$-phase (770 - 1100~K). This indicates a very strong anharmonicity of the phonon dynamics in this phase. The phonon 
modes are most affected by the transition at  T$_C$ (1050 - 1100 K) where BFO changes from the polar ferroelectric $\alpha$-phase to a paraelectric intermediate $\beta$-phase. This holds 
in particular for the low-frequency part of the spectrum and makes the statement that these modes could provide coupling of magnetism and ferroelectricity all the more plausible. In the 
$\beta$ phase (1100 - 1170 K), the intermediate frequency part is subject to a considerable downshift resulting in a complete filling of the gap that was separating the intermediate and 
low-energy regions. The system undergoes a second phase transition around 1200~K to reach a third structural phase ($\gamma$) as clearly reflected in the phonon spectra. The weight of the 
low-energy phonon modes diminishes when compared to the $\alpha$ and $\beta$ phases in favor of two new strong features at 20 and 35 meV. These phase transitions unravel changes across 
various ferroic order parameters that are separated over a temperature range. As ferroelectricity and magnetism in BFO are driven by the Bi and Fe sites, respectively, the partial PDOS 
reveal clearly the frequency ranges where they contribute. Interestingly, the ferroelectic-to-paraelectric phase transition involves large changes in the dynamics of the 
Bismuth cations in terms of vanishment of the related off-centered displacements. The abrupt change observed in both phase transitions ($\alpha$ $\rightarrow$ $\beta$ and 
$\beta$ $\rightarrow$ $\gamma$) confirm that they are of a first order nature. These results provide a reliable support to other works using different 
techniques~\cite{pnd1,pnd2,finittemp}.

As already mentioned, the oxygen contribution to the observed phonon spectra changes considerably upon heating. It is fully smeared out and concomitantly lost intensity in the $\gamma$-phase. This is 
indicative of a weakening of Fe-O bonding leading finally to bond breaking with the consequence of oxygen release and vacancy creation. This result is quite interesting since 
it provides a strong argument for the occurrence of the high-spin to low-spin transition mentioned above, and which is a consequence of the reduction of Fe$^{3+}$ to Fe$^{2+}$ to maintain 
the charge balance due to the presence of oxygen vacancies~\cite{pnd1,pnd2,revphysandapps}. Similar dynamic effects have been seen in superionic oxygen conductors with an open oxygen 
network structure~\cite{Paulus-JACS}. The newly visible vibrational features in the $\gamma$ phase at 20 and 35 meV could be due to an increase in cell volume after Fe reduction 
(Fe$^{3+}$ $\rightarrow$ Fe$^{2+}$) which results in a red shift of closely similar modes observed in the $\alpha$ phase at 25 and 40 meV, respectively. 

In summary we have carried out inelastic neutron scattering measurements of the phonon spectra of BiFeO$_3$ which is a unique model system and multiferroic material with multiple order 
parameters. We have found a strong temperature evolution that indicates both strong anharmonicity and subtle dependence of phonons on structural details. The phonon modes are most 
affected by the transition at the Curie temperature T$_C$. Three phases $\alpha$, $\beta$ and $\gamma$, in agreement with and in support of other works~\cite{pnd1,pnd2}, are highlighted 
by their clearly different vibrational signatures. Ab-initio calculations provide a phonon density of states in good agreement with the observations. The atomistic vibrational 
understanding extracted from these calculations allows gaining reliable insights into the dynamic behavior of BFO upon temperature variation. On the electronic side the occurrence of a 
high-temperature metal-insulator and Fe$^{3+}$ $\rightarrow$ Fe$^{2+}$ spin transitions, previously inferred via diffraction experiments~\cite{pnd1,pnd2},
looks very plausible and the present results are consistent with these findings. The high temperature neutron diffraction~\cite{pnd1,pnd2} and
optical data~\cite{massa} have been associated with spin changes and also decomposition and our results reveal large changes in the phonon density of states, consistent with these
findings. We estimate that these results are of considerable relevance for the field of multiferroism in general and for multiferroism in this specific material, in
particular. They provide valuable arguments to better interpret and support results obtained with other techniques~\cite{massa,pnd1,pnd2,bfo5,irphonons,phasestabstructtemp,finittemp,revphysandapps,calcmagnetoelec}.
\newpage
%
\newpage
%

\begin{thebibliography}{30}
\bibitem[{\citenamefont{28}}]{furnace}
\bibinfo{title}{A dedicated high-temperature sample environment was used for the reported inelastic neutron scattering measurements. Due to a leakage current issue, data at 300 K was not collected and the present interest is focused on the evolution of the system in the high-temperature range (500 - 1200 K) covering both the T$_N$ $\sim$ 600 - 650~K and T$_C$ $\sim$ 1050 - 1100~K.}
\end{thebibliography}
\end{document}